\documentclass{elsarticle}

\usepackage[margin=2.5cm]{geometry}

\usepackage{lineno}
\usepackage[colorlinks=true,
           urlcolor=blue,
           citecolor=blue,
           linkcolor=blue,
           bookmarks=false,
           bookmarksnumbered,% For PDF navigation
           linktocpage=true
           ]{hyperref}

\usepackage{color}

\modulolinenumbers[5]

\journal{Future Generation Computer Systems}

\newif\iffinal

\finaltrue

\definecolor{olive}{rgb}{0.5, 0.5, 0.0}
\iffinal
  \newcommand\change[1]{}
  \newcommand\kyle[1]{}
  \newcommand\victoria[1]{}
  \newcommand\kacper[1]{}
  \newcommand\mihael[1]{}
\else
  \newcommand\change[1]{{\color{blue}[Change: #1]}}
  \newcommand\kyle[1]{{\color{blue}[Kyle: #1]}}
  \newcommand\victoria[1]{{\color{red}[Victoria: #1]}}
  \newcommand\kacper[1]{{\color{green}[Kacper: #1]}}
  \newcommand\mihael[1]{{\color{olive}[Mihael: #1]}}
\fi

\newcommand{\wt}{Whole\,Tale}

\begin{document}

\begin{frontmatter}

%Capturing the ``\wt\ '' of Computational Claims: 
\title{\textbf{Computing Environments for Reproducibility:\\Capturing the ``\wt''}}
% \tnotetext[mytitlenote]{Fully documented templates are available in the elsarticle package on \href{http://www.ctan.org/tex-archive/macros/latex/contrib/elsarticle}{CTAN}.}

%% Group authors per affiliation:
%\author{Bertram Ludaescher\fnref{myfootnote}}
%\address{School of Information Sciences, University of Illinois at Urbana-Champaign}
%\fntext[myfootnote]{School of Information Sciences, University of Illinois at Urbana-Champaign}

%% or include affiliations in footnotes:
%\author[mymainaddress,mysecondaryaddress]{Elsevier Inc}
%\ead[url]{www.elsevier.com}

\author[address5]{Adam Brinckman}
\author[address2]{Kyle Chard\corref{mycorrespondingauthor}}
\cortext[mycorrespondingauthor]{Corresponding author}
\ead{chard@uchicago.edu}
\author[address3]{Niall Gaffney}
\author[address2]{Mihael Hategan}
\author[address4]{Matthew B. Jones}
\author[address6]{Kacper Kowalik}
\author[address3]{Sivakumar Kulasekaran}
\author[address1,address6]{Bertram Lud\"ascher}
\author[address4]{Bryce D. Mecum}
\author[address5]{Jarek Nabrzyski}
\author[address1,address6]{Victoria Stodden}
\author[address5,address7]{Ian J.~Taylor}
\author[address1,address6]{Matthew J.~Turk}
\author[address6]{Kandace Turner}  
\address[address1]{School of Information Sciences, University of Illinois at Urbana-Champaign}
\address[address2]{Computation Institute, University of Chicago and Argonne National Laboratory}
\address[address3]{Texas Advanced Computing Center, University of Texas at Austin}
\address[address4]{National Center for Ecological Analysis and Synthesis, University of California at Santa Barbara}
\address[address5]{Center for Research Computing, University of Notre Dame}
\address[address6]{National Center for Supercomputing Applications, University of Illinois at Urbana-Champaign}
\address[address7]{School of Computer Science, Cardiff University, Cardiff, UK}

\begin{abstract}
The act of sharing scientific knowledge is rapidly evolving away from traditional articles and presentations to the delivery of executable objects that integrate the data and computational details (e.g., scripts and workflows) upon which the findings rely. This envisioned coupling of data 
and process is essential to advancing science but faces 
% BL: removed "deep" (or can we clarify which barriers, i.e., technical or institutional are particularly deep? It seems we're OK on the technical side at least..
technical and institutional barriers. The \wt\ project aims to address these barriers by connecting computational, data-intensive research 
efforts with the larger research process---transforming the knowledge discovery 
and dissemination process into one where data products are united with research articles 
to create ``living publications'' or \emph{tales}. The \wt\ focuses on the full spectrum of science, empowering users in the long tail of science, and power users with demands for access to big data and compute resources. We report here on the design, architecture, and implementation of the \wt\ environment. 
\end{abstract}

\begin{keyword}
\texttt{living publications \sep reproducibility \sep provenance  \sep data sharing \sep code sharing}
%\MSC[2010] 00-01\sep  99-00
\end{keyword}

\end{frontmatter}

%\linenumbers

\section{Introduction}\label{sec:intro}

The pervasive use of computation for scientific discovery has ushered in a new type of scientific research process. Researchers, irrespective of scientific domain, routinely rely on large amounts of data, specialized computational infrastructure, 
and sophisticated analysis processes from which to test hypotheses and derive
results. While scholarly research has evolved significantly over the past decade, 
the same cannot be said for the methods by which research processes are captured and disseminated. 
In fact, the primary method for dissemination---the scholarly publication---is largely unchanged since the advent of the scientific journal in the 1660’s. This disparity has led many to
argue that the scholarly publication is no longer sufficient to verify, reproduce, 
and extend scientific results~\cite{peng11reproducible, kratz14datapub, alsheikh11public,stodden_setting_2012,donoho_reproducible_2009,stodden_implementing_2014}.
%stodden_resolving_2013}.

The challenges associated with rethinking the scholarly publication model are complicated by
the pervasive increase in the collection and analysis of data, coupled with dramatic increases 
in computational power, and new methods for investigation such as data-driven discovery. 
% These factors have created new avenues for investigation---from 
% small to massive datasets, and from small scale computation to state-of-the-art 
% high-performance computing.
The scientific landscape is now littered with a vast array of powerful
cyberinfrastructure for acquiring, storing, analyzing, publishing, and archiving data. However, current approaches regarding the dissemination, validation, and verification of computationally based 
research outcomes do not yet accommodate this reality. 
Despite the increasing recognition of the need to share all aspects of the research process, 
scholarly publications today are often \emph{disconnected} from the underlying data and code 
that produced the findings. While efforts have been made to support data publication~\cite{dataverse,figshare,chard15publication}, 
\emph{``unfortunately, the vast majority of data submitted along with 
publications are in formats and forms of storage that makes discovery and reuse difficult or 
impossible''}~\cite{copdess}. Studies of published data have also shown
that data availability decays with time, if the data are available at all~\cite{views14decline}.

To address these challenges we present \wt~\cite{ludascher16wt}, a research environment that captures and, at the time of publication, exposes salient details of the entire research process via access to persistent versions of the data and code used, workflow provenance, and data lineage (including parameter settings, intermediate, and output data). The \wt\ directly addresses the transformation of the scientific enterprise to deeply computational research by supporting the entire 
research pipeline, from pre-publication collaboration, through publication, and to post-publication 
access and re-use in the broader scientific community.
We present here the design and current implementation of the \wt\ environment (\url{https://dashboard.wholetale.org}).

%As a key step towards our vision, we propose a series of integration tools and methods with demonstrated applications. Traditionally, the first layer of scholarly output has been accomplished through the production and dissemination of research articles. As data become more open and transportable, a second layer of research output, linking publications to the associated data, is emerging. This is now rapidly followed by the recognition of an important and new third layer: communicating the process of inquiry itself, i.e., a complete computational narrative, through the linking and sharing of methods, source code, and data, thereby introducing a new era of reproducible science and accelerated knowledge discovery.

The \wt\ strengthens the three layers of scholarly publication: scholarly process, data, 
and computational analysis. Traditionally, the first layer of scholarly publication has been accomplished 
through the production and dissemination of research articles. As data become more open and transportable, a 
second layer of research output, linking publications to the associated data, has emerged~\cite{kratz14datapub}. This is now  followed by the recognition of an important and new third layer: communicating the process of inquiry itself, 
i.e., a complete \emph{computational narrative}, through the linking and sharing of methods, source code, and data, 
thereby introducing a new model of reproducible science and accelerated knowledge discovery~\cite{Stodden1240}. The \wt\ strengthens the second layer (linking data, code, and digital scholarly objects to publications) and also builds
a robust third layer that integrates all parts of the research story into a computational \emph{tale} (conveying the holistic experience of reproducible scientific inquiry, i.e., sharing the source code, data, and methods, along with the computational environment in which inquiry is conducted) and making both layers accessible from the scholarly publication. To the user it thus appears that \emph{by sharing a paper as a tale, the narrative is shared together with an on-demand, virtual computer that is preloaded with all the relevant data, methods, software packages, and analysis frontends needed to reproduce, tinker with, or even extend the paper}. 

%There is frequently no way to trace findings in publications back through the
%originating computations and data. We propose to remedy this gap in two ways: 1) integrate existing
%cyberinfrastructure that supports the entire computational process underlying discovery, thus
%simplifying the ability for researchers to conduct computational researcher; and 2) and capturing and
%delivering relevant workflow and processing provenance that will be discoverable and
%accessible from the associated publication. In addition, \wt\ will be a collaborative
%environment where data providers, application developers, and data consumers collaborate and
%create end-to-end workflows converting data to information using reproducible computational methods.

The \wt\ environment can also be seen as a form of \emph{science gateway}~\cite{wilkins07gateways}: it 
simplifies access to a vast array of cyberinfrastructure for a broad range of domain scientists. The architecture
described herein builds upon advances made within the science gateways community to leverage external
services for core functionality such as user authentication and authorization, data management, and 
 management of computational resources. Finally, the \wt\ architecture is based on extensible APIs that 
can be leveraged by other gateways for recording computational processes, importing 
and managing data, issuing identifiers, and sharing and publishing reproducible tales.

The remainder of this article is structured as follows. In Section~\ref{sec:science} we first
present examples from three scientific domains that highlight some of the challenges commonly faced by computational scientists. In Section~\ref{sec:reproducibility} we describe prior and current efforts towards enabling reproducible research. We then describe high level requirements of the \wt\ in Section~\ref{sec:design} before presenting
the architecture and current implementation in Section~\ref{sec:architecture}.
In Section~\ref{sec:relatedwork} we review related work. Finally, we summarize our contributions
in Section~\ref{sec:summary}.

\section{Science Narratives}\label{sec:science}

% In future it would be awesome if we could link to a tale for each of our examples.

We first describe three scientific domains that, like many others, have
embraced computational and data-driven science. We 
focus specifically on examples that elucidate usage requirements for the \wt. 
% including
% data and computational requirements, analysis processes, and reproducibility requirements. 

\subsection{Materials Science}
Materials scientists are now generating vast amounts of computational and 
experimental data from a wide set of user facilities (e.g., the APS, SNS, NSLS-II), 
from simulations at Leadership Computing Facilities, from individual 
research labs, and from high-throughput experiments. To address the deluge
of high quality data there are now numerous data repositories designed to 
store and provide access to curated materials data including: the Materials Data Facility (MDF)~\cite{blaiszik16mdf}, 
Materials Project~\cite{jain13materialsproject}, Citrination~\cite{omara16citrination}, and NoMaD (Novel Materials Design)~\cite{thygesen16nomad}. With these rich materials 
data sources, opportunities are available to
conduct new types of analysis and for researchers to supplement their own data to expand investigations.
 
Computational approaches are having a profound effect on materials science. 
Over the past several decades, concurrent advancements in physics-based simulation methods and 
computing power have made it possible to model material behavior on a large range of length and
time scales~\cite{yip07handbook}. Consequently, computational tools are starting to become an integral part
of designing new materials~\cite{council04accelerating}. For example, quantum-mechanics-based calculation tools 
are routinely used in the development of structural metals and semiconductors, among other materials~\cite{curtarolo13high}. 
Increasingly, these computational processes are based on new machine
learning methods to construct rich models of materials properties~\cite{hill16materials, ward16materials, rajan05materials}. 
With such changes, researchers are increasingly in need of new methods
for publishing not only references to the data used to derive results
but also the models and computational processes that underpin results. 
 
As one example of these changes we describe the process undertaken by Ward
et al. to design new metallic glasses using machine learning models~\cite{ward16materials}. The authors first assembled a collection of materials data~\cite{kawazoe97phase}. In order to build 
a machine learning model, they used custom software to transform the raw data, text strings describing each material’s composition and properties, to a form compatible with their models: finite length vectors of physically-meaningful inputs. They trained machine learning models using Weka~\cite{hall09weka} and employed the models to scan over several million compositions to identify novel glass-forming alloys. 
% The authors also performed several cross-validation experiments to validate the effectiveness of their model. 
In an effort to make their methods verifiable and reproducible the authors published their workflows (as text input and data files) as supplementary information to the paper. Using \wt\ these researchers could streamline this process via access to a large and varied amount of data, a platform for conducting their analyses
using containerized frontends, and the ability to subsequently publish their entire method (including data and analyses) with a persistent identifier. Readers of their
manuscript could then view their methods (as a tale) and reproduce the
exact steps taken within the \wt\ environment.

\subsection{Astronomy}

Detailed analysis and visualization of astronomical datasets, particularly those generated from computational simulations, requires both access to the original underlying data (or catalogs of reduced data products) and access to computing resources.  One particular example being explored by the \wt\ project is that of studying the formation of the first stars and galaxies in the universe (see, for instance,~\cite{2015MNRAS.452.2822S}), but another common use case is that of galaxy formation~\citep{2014ApJS..210...14K}.  These simulations are conducted on large-scale computing resources; typically, the analysis utilizes community packages such as \texttt{yt}~\citep{2011ApJS..192....9T} and, through the development of scripts and interactive analysis sessions, produces either publication-level plots or reduced data products that can be reanalyzed at a later date.  In the case of observational astronomy, multiple datasets may need to be synthesized to create a unified understanding of either a particular class of object or a region of the sky; in many cases, this will require small ``slices'' of data from many different sources (e.g., from several public registries) to be combined.

For the specific case of analysis and visualization of simulations, \wt\ will provide access to a collaborative environment, where scripts and analysis methods can not only be transplanted seamlessly between datasets, but where they can be collaborated on between individuals -- such as an advisor and a student.  Researchers will be able to conduct simulations, make available the results of those simulations inside the \wt\ environment, and then conduct their analysis in that environment directly.  The scripts that produce plots and analysis products for publication purposes will be combined with these datasets to form a tale, which can then be accessed, remixed, and modified for subsequent analysis either by those researchers or by others.  This will open up new avenues for discovery, which at present is constrained both by the difficulty of providing access to data and by the difficulties inherent in collaborating on the specific methods of analysis and visualization.

\subsection{Archaeology}

A set of grand challenges in archaeology\footnote{We adopt the spelling of the Society for American Archaeology \cite{little06}.}
 have been identified by the community through a crowd-sourced effort and synthesis workshop \cite{kintigh2014granda}. While archaeological data and research are essential to addressing fundamental questions, e.g., about the origin and trajectories of civilizations or the response of societies to climate change, the community lacks the capacity for acquiring, managing, analyzing, and synthesizing datasets needed to address such important questions. This in turn led to recommendations for computational infrastructure, tools, and scientific case studies to demonstrate archaeology's ability to contribute to transdisciplinary research on long-term social dynamics~\cite{kintigh2015cultural}. 

One such project is SKOPE~\cite{skope}, which is developing an online resource and toolkit for paleoenvironmental data and models that will enable researchers to easily discover, explore, visualize, and synthesize knowledge of environmental factors most relevant to humans in the past. SKOPE's focus on transparent, reproducible research, facilitated by different forms of provenance, makes it an ideal partner project and science driver for \wt.
% For example, in connection with the \wt\ science working group focusing on computational archaeology, we are offering a 2017 summer internship project that will focus on ``telling the whole tale behind a climate reconstruction based on PaleoCAR", i.e., which will create integrated representations of various dimensions of provenance of the results of a particular scientific study \cite{bocinsky2014000year}. 
To address the complex, multi-stage workflows inherent in this domain, these researchers will employ YesWorkflow \cite{mcphillips_yesworkflow:_2015,mcphillips2015retrospective} to create graphical, queryable representations of each of the computational workflows enacted as part of the research. These workflows capture the \emph{prospective} provenance of all data products generated during the study. Such workflows can be used within the \wt\ environment to create hybrid forms of provenance \cite{pimentel2016yin,zhang17revealing} that combine prospective with retrospective provenance information of intermediate and final data products, complete with records of the specific program executions involved, the values of program arguments applied, and---where possible---the values of key variables within the programs themselves as exposed by YesWorkflow.

Finally, there are several community efforts such as ``How To Do Archaeological Science Using R''~\cite{archR} that aim to improve community practice: i.e., instead of sharing methods only via traditional  publications, reproducibility and reuse are facilitated by authors communicating their methods also via open code repositories and using tools to package computational narratives as research compendia for R \cite{marwick2017packaging,bocinsky17provathon}. These efforts provide current, real-world science use cases that \wt\ aims to support and enhance.

\section{Toward Reproducibility}\label{sec:reproducibility}

As we noted at the outset, the complexity and details of the computational steps that gave rise to the scientific conclusions is typically impossible to capture in a traditional publication~\cite{victoria01}. However, increasing the transparency of computational findings falls on several stakeholders. As researchers move toward greater reproducibility it is essential that funding agencies, publishers, and local incentives align to support this transition. Steps are being taken at all stakeholder levels, yet open questions remain. % BL: Such as?

Researchers were the first to implement new practices that encompassed reproducibility in computational research. The earliest to our knowledge was the introduction of ``really reproducible research'' in 1992 by the Stanford Exploration Project~\cite{victoria02} which introduced reproducibility standards for electronic documents that contain computational results. Most recently~\cite{victoria03} exhorted the community to ensure that the digital artifacts needed for verification (i.e. data, code, workflows) are made available to the community in a usable form with the publication. The recommendations were:

\begin{enumerate}
	\item To facilitate reproducibility, share the data, software, workflows, and details of the computational environment in open trusted repositories.
    \item To enable discoverability, persistent links should appear in the published article and include a permanent identifier for data, code, and digital artifacts upon which the results depend.
    \item To enable credit for shared digital scholarly objects, citation should be standard practice~\cite{victoria04,victoria05}.
    \item To facilitate reuse, adequately document digital scholarly artifacts.
    \item Journals should conduct a Reproducibility Check as part of the publication process and enact the Transparency and Openness Promotion (TOP) Standards at level 2 or 3~\cite{victoria06}.
    \item Use Open Licensing when publishing digital scholarly objects~\cite{victoria07,victoria08}.
    \item To better enable reproducibility across the scientific enterprise, funding agencies should instigate new research programs and pilot studies.
\end{enumerate}

% Recommendation 5 references the TOP, or Transparency and Openness Promotion, Guidelines. 
To date, more than 5000 journals have signed on to the TOP Guidelines~\cite{nosek1422,victoria07}. 
Journals are progressively taking steps to encourage the submission and publication of reproducible computational research~\cite{victoria10}. 
%With the exception of Recommendation 5, responsibility for the implementation of the above recommendations rests with funding agencies, journals and scientific societies, and institutions and promotion and tenure committees. 

Funding agencies are also moving toward implementing reproducible research. The National Science Foundation requires the disclosure of data and software created in the course of research they fund. The NSF Award \& Administration Guide (AAG) Chapter VI.D.4. (October 2016) reads:

\begin{itemize}
\item b.	Investigators are expected to share with other researchers, at no more than incremental cost and within a reasonable time, the primary data, samples, physical collections and other supporting materials created or gathered in the course of work under NSF grants. Grantees are expected to encourage and facilitate such sharing. [...]

\item c.	Investigators and grantees are encouraged to share software and inventions created under the grant or otherwise make them or their products widely available and usable.

\end{itemize}

The NSF held an agency-wide Director's Symposium ``Robust and Reliable Science: The Path Forward'' on September 10, 2015. More recently, on February 25-26, 2017, the National Science Foundation's Directorate on Mathematical and Physical Sciences held a workshop ``Systematic Approaches to Robustness, Reliability, and Reproducibility in Scientific Research'' fomenting a discussion around reproducibility~\cite{victoria11}. In December of 2016 the Advisory Committee to the Computer and Information Science and Engineering Directorate at NSF released a report ``Realizing the Potential of Data Science''~\cite{victoria12} which included recommendations on reproducibility:

\begin{itemize}
\item Recommendation 2: Invest in research into data science infrastructure that furthers effective data sharing, data use, and life cycle management: ... Research outcomes should ultimately be translatable to infrastructure that enables access to data in ways that: ... (iii) support reproducibility; (iv) support access, provenance, sustainability, and other life cycle challenges.
\item Recommendation 3: Support research into effective reproducibility: Develop research programs that support computational reproducibility and computationally-enabled discovery, as well as cyberinfrastructure that supports reproducibility.
\end{itemize}

Scientific societies, in part in their role as publishers, are taking steps toward reproducibility as well. The ACM has implemented a system of badging for publications that have digital artifacts available~\cite{victoria13}. In November of 2016 IEEE held a workshop on publication practices for reproducibility, ``The Future of Research Curation and Research Reproducibility''~\cite{victoria14}.

Finally, the National Academies of Sciences, Engineering, and Medicine, released a report in April 2017, Fostering Integrity in Research~\cite{NAP21896}, which contained two recommendations regarding reproducible research:

\begin{itemize}

\item Recommendation 6: Through their policies and through the development of supporting infrastructure, research sponsors and science, engineering, technology, and medical journal and book publishers should ensure that information sufficient for a person knowledgeable about the field and its techniques to reproduce reported results is made available at the time of publication or as soon as possible after publication. 
\item Recommendation 7: Federal funding agencies and other research sponsors should allocate sufficient funds to enable the long-term storage, archiving, and access of datasets and code necessary for the replication of published findings. 

\end{itemize}

There have been concurrent advances in European open access and open data policy. In 2003 the Berlin Declaration on Open Access to Knowledge in the Sciences and Humanities was signed by nearly 300 stakeholder groups including research and educational institutions, libraries, museums, funding agencies, and governments from around the world to help establish the Internet as the primary medium of communication and dissemination of scientific knowledge~\cite{berlindeclaration}.  
EUDAT~\cite{eudat}, a European-based effort to share and preserve data across international border and across research disciplines was started in 2012 and continues actively today.
OpenAIRE~\cite{openaire} is a European repository effort to, in part, link data to publications and was started in 2009. 
On the infrastructure side, EuroCloud~\cite{eurocloud} was launched in 2010 in part to support cloud based research and innovation in Europe.

The 2017 version of the European Code of Conduct for Research Integrity explicitly mentions data integrity~\cite{eu-code}. 
Their list of ``Good Research Practices" includes:

\begin{itemize}
\item Research institutions and organisations
support proper infrastructure for the
management and protection of data
and research materials in all their forms
(encompassing qualitative and quantitative
data, protocols, processes, other research
artefacts and associated metadata) that are
necessary for reproducibility, traceability
and accountability.
\end{itemize}

% The 2017 Code of Conduct further states, under "Data Practices and Management":

% \begin{itemize}
% \item Researchers, research institutions and
% organisations ensure appropriate stewardship 
% and curation of all data and research materials,
% including unpublished ones, with secure
% preservation for a reasonable period.
% \item Researchers, research institutions and
% organisations ensure access to data is as
% open as possible, as closed as necessary,
% and where appropriate in line with the
% FAIR Principles (Findable, Accessible,
% Interoperable and Re-usable) for data
% management.
% \item Researchers, research institutions and
% organisations provide transparency about
% how to access or make use of their data and
% research materials.
% \item Researchers, research institutions and
% organisations acknowledge data as legitimate
% and citable products of research.
% \item Researchers, research institutions and
% organisations ensure that any contracts or
% agreements relating to research outputs
% include equitable and fair provision for the
% management of their use, ownership, and/or
% their protection under intellectual property
% rights.
% \end{itemize}

\noindent A Dagstuhl seminar on \emph{Reproducibility of Data-Oriented Experiments} \cite {freire2016reproducibility} summarizes:
\begin{itemize}
\item
Transparency, openness, and reproducibility are vital features of science. Scientists embrace these features as disciplinary norms and values, and it follows that they should be integrated into daily research activities. These practices give confidence in the work; help research as a whole to be conducted at a higher standard and be undertaken more efficiently; provide verifiability and falsifiability; and encourage a community of mutual cooperation. They also lead to a valuable form of paper, namely, reports on evaluation and reproduction of prior work. Outcomes that others can build upon and use for their own research, whether a theoretical construct or a reproducible experimental result, form a foundation on which science can progress. Papers that are structured and presented in a manner that facilitates and encourages such post-publication evaluations benefit from increased impact, recognition, and citation rates.
Experience in computing research has demonstrated that a range of straightforward mechanisms can be employed to encourage authors to produce reproducible work. These include: requiring an explicit commitment to an intended level of provision of reproducible materials as a routine part of each paper's structure; requiring a detailed methods section; separating the refereeing of the paper's scientific contribution and its technical process; and explicitly encouraging the creation and reuse of open resources (data, code, or both).
\end{itemize}

As noted in several of the recommendations discussed above, new research and new technologies are needed to implement reproducible computational research, and the \wt\ represents one initiative to address these gaps in our research and dissemination infrastructure. 

% Better in RW? 
% Other projects exist to help close similar gaps in a variety of areas. For example, the journal Image Processing Online (\url{http://ipol.im}) provides reproducible publications for the image processing community, Code Ocean provides reproducibility functionality for IEEE publications, the Madagascar project extends the reproducibility functionality described by Claerbout and Karrenbach in 1992 (\url{http://www.ahay.org}), and the WaveLab project pioneered reproducibility in signal processing (\url{http://statweb.stanford.edu/~wavelab/}), just to name a few.

\section{Design Requirements}\label{sec:design}

The \wt\ project is intended to support the lifecycle of data. This means that all parts of the lifecycle, from data ingest or creation through to publication of the resulting scholarly objects such as data, code, workflows, and manuscripts, should be manageable within the \wt\ framework. Our discussion of design and implementation therefore reflects an integrated view of the generation of computational scientific findings that includes all these research activities. This integrated approach to research is crucial to enable reproducibility and downstream re-use of scholarly objects. 

% [delete this since I largely encapsulated it above] By tying together remote storage, access to resources, and publishing, while removing barriers to combining, processing, and analyzing data, the  \wt\ environment aims to provide a researcher-oriented system for the full lifecycle of scientific inquiry, from experimental design through to publication. 

To provide such support, \wt\ incorporates data ingestion, 
identity management, data publication, and the deployment of user-facing ``frontends.''  We use the term \emph{frontend} to describe any environment in which data can be operated on, ranging from
terminals with a command-line interface to specialized analysis programs. 
Examples of common frontends include interactive notebooks (e.g., Jupyter, RStudio),
HTML5 web-apps, and domain-specific GUIs (e.g., OpenRefine).
We briefly describe the requirements
in several core areas to motivate the architecture presented in the following section.

% KC removed as this figure is fairly dated
% The high level workflow for using the Whole
% tale environment is shown in Figure~\ref{fig:workflow}.

% \begin{figure}[ht!]
% \centering
%   \includegraphics[trim=0in 0in 0in 0in,clip,width=0.75\columnwidth]{workflow.png}
% \caption{\wt\ workflow. The scientist's frontend can seamlessly access research data and carry out analyses in the \wt\ environment. Digital research objects, such as code, scripts, or data produced during the research, are shared between collaborators. When the researcher is ready, these are bundled with the paper to produce a tale that is accessible and can be persisted.  When the tale is published in the scholarly record, it is augmented with embedded links to data and code (with DOIs) used in the production of the scientific results.\label{fig:workflow}}
% \end{figure}

\subsection{Data Ingestion}
\label{sec:data_ingestion}
% There is an ever growing collection of domain-specific, institution, project, 
% and publisher data repositories. 
Researchers now have access to an enormous amount of data from 
sources such as data repositories, instruments, and local storage. 
% large collection
% of data repositories, the
% variety of instruments, computers, and other sources from which 
% data are derived, and the many temporary and permanent storage locations in which data are stored. 
Researchers who want to act upon data, for example to test a hypothesis
or reproduce a result, must first discover and then obtain access to
data distributed across many possible locations. The \wt\ environment aims to 
reduce these barriers by providing mechanisms by which researchers
can ingest data from a wide variety of sources. We focus initially
on four commonly used data sources: 

\begin{itemize}
	\item \textbf{Data Repositories:} 
    There is an increasing number of domain-specific, institutional, project-centric, and publisher-owned data repositories. 
    Many data repositories, including the Materials Data Facility (MDF) and DataONE~\cite{michener15dataone}(a federation of repositories), support common interfaces for accessing published metadata and data.
    These interfaces include the 
    Open Archives Initiative Protocol for Metadata Harvesting (OAI-PMH)~\cite{oaipmh}
    as well as many custom REST interfaces.

    \item \textbf{Storage systems:} Research data is distributed across a range
    of local systems, from instruments to archival storage. Each storage system
    implements one of many interfaces for accessing that data (e.g., object storage,
    tape interfaces, cloud storage, high performance file systems, etc.)  
    
    \item \textbf{Web accessible data:} There is a vast amount of data stored
    on web pages or web-based data repositories. In these cases, data can be discovered and downloaded using HTTP-based tools.

    \item \textbf{Local data:} Much research data exists on researchers' personal
    computers, shared clusters, or otherwise inaccessible (in terms of a common API) devices. 
\end{itemize}

\subsection{Analysis Frontends}

% The number of analysis and compute environments used by researchers is
% limitless. 
% We use the term ``frontend'' to describe any environment in which
% data can be operated on, ranging from terminals to specialized analysis programs. 
% Examples of common frontends include 
% interactive notebooks, HTML5 web-apps, and domain-specific GUIs. 
As mentioned above, we use the term \emph{frontend} to describe any environment in which
data is operated on, ranging from command-line terminals to specialized analysis programs. The choice of frontend used for a specific scientific analysis
may be based on analysis requirements, data type, or user preference.
One example frontend that is commonly used by researchers is the Jupyter notebook environment~\cite{jupyter}. 
Jupyter notebooks support multiple language backends (Python, R, Julia, and many others), widget development for interactive exploration, file editing, and shell activity within a unified, web-based environment. 
% Jupyter notebooks provide enormous
% flexibility, enabling very deep inquiry, to the level of direct file access, and is thus both 
% valuable for deployment as well as a ``pathfinder'' for specialized environments serving smaller, more focused communities of users.

To address the needs of a wide range of users and use cases, the \wt\ must support an extensible set of frontends. Users coming to \wt\ should be able to search available frontends by the types of data they can support,  the user interface offered (web, command-line, digital notebook, etc.), and which user provided them. 
Having discovered a frontend, users should then
be able to rapidly deploy them on-demand and access data
directly from within the frontend. %, and extend the research in new ways.
The \wt\ environment must manage the execution of a frontend while
also capturing the steps followed by a user such that the entire frontend
can be packaged, published, and shared with others.

\subsection{Persistent Identification} 
One of the primary goals of the \wt\ is to enable publication and identification
of scholarly objects, where the term scholarly object is used to describe more than a traditional publication but also
data and computational processes. A flexible identification and resolution service
is required to allow persistent identifiers (e.g., DOIs, ARKs, Handles) to be associated with these
different objects. 
% It is crucial that the referenced objects, that live within the Whole
% tale environment, can be accessed externally, using standard protocols.  
Furthermore, models are needed to allow
researchers to organize their objects in different ways, both for their own purposes and
also to simplify collaboration and discovery. As such, the \wt\ must provide a way for researchers to organize their scholarly objects. 
% Thus enabling users and communities to develop their own namespaces for datasets 
% that exist locally or remotely so that 
% semantically-meaningful names can be provided.

\subsection{Authentication and Authorization}
% Furthermore, many data repositories
% require knowledge of one's identity to provide authorized access to data. 
The \wt\ aims not to reinvent existing capabilities but rather
to interoperate with existing services
and cyberinfrastructure providers
(e.g., repositories, compute environments, libraries). Each of these existing providers
might be managed independently with proprietary identities and authorization models. 
It is therefore important that \wt\ adhere to existing authentication
models and ensure that digital artifacts accessed and created during the exploration 
and publication of scholarly objects are correctly authorized. 
% requires an overarching system that ties together that person's identity with 
% the actions they conduct in the environment.  

When designing the authentication model it is desirable that researchers 
are able to sign in once, 
access a range of supported services, and have their identity and permissions 
be used securely across services. 
Given the rate of identity proliferation, the authentication system should allow
researchers to authenticate with their preferred identity (e.g., campus identity, 
ORCID, Google account)
and control authorization at a fine grained level (e.g., revoking access
when needed). 
Rather than restrict the identities used in the system, we instead associate identities with actions and artifacts, and allow other users to determine 
trust based on knowledge of the identity used.
To enable extensibility to new tools and services, \wt\ must support standard authentication
and authorization protocols through which external services and clients can easily integrate
with the system.
The \wt\ focuses on acting upon research data, we therefore consider 
issues related to sensitive data (e.g., restricted medical or government data) outside
the scope of this work.

\subsection{Reproducibility: Defining a Tale}

The final, and perhaps most important, aim of the \wt\ is to 
define a model for reproducibility by capturing the data, methods, metadata, and provenance of a particular research activity within the system. We refer to this entity as
a \emph{tale}. As has been observed time and again, successful adoption of 
new models is often related to the ease by which they can be used. As such, 
it is crucial that capturing, publishing, and ``replaying'' a tale is simple and unobtrusive: the relevant
provenance of an analysis should be transparently recorded 
without requiring users to manage or record the data and computational
process used in their work.
Having created one or more tales, researchers should be able to simply share
them with others, publish them to connected repositories, associate
persistent identifiers, and link them to publications. Other researchers
who access a tale should, just as simply, be able to instantiate a version
of the tale and execute it in the same state as it was when published. Tales also contain Intellectual Property metadata with licensing information for its components (data, scripts, workflow information, etc), which is crucial to enabling ease of re-use, reproducibility, and broad access.

\section{Architecture and Implementation}\label{sec:architecture}

\begin{figure}[ht!]
\centering
  \includegraphics[trim=0in 0in 0in 0in,clip,width=.75\columnwidth]{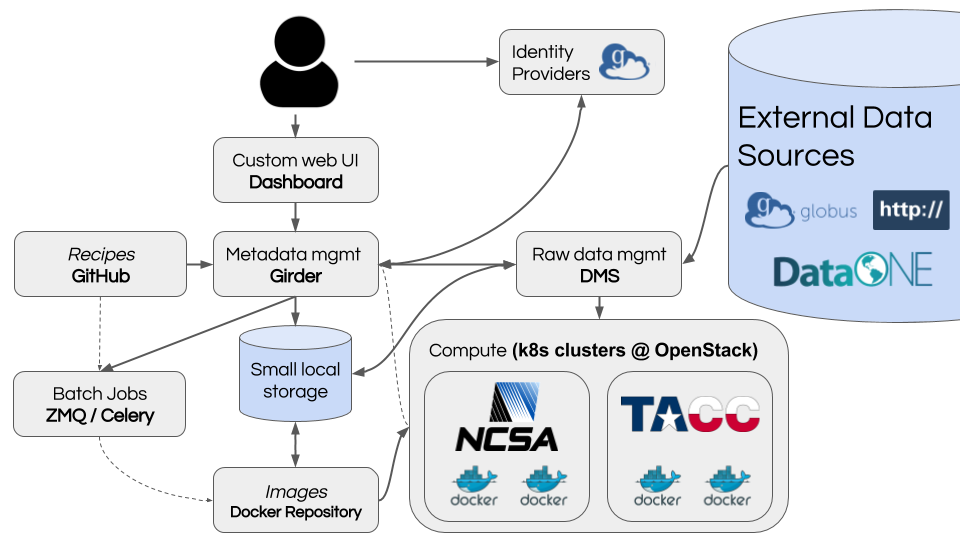}
\caption{\wt\ architecture. Users connect via common web interfaces. Microservices manage data, users, frontends, and tales. The execution environment manages the execution of frontends in containers on different compute resources. \label{fig:arch}}
 \end{figure}

The \wt\ architecture uses a range of flexible APIs to enable users to ingest and manage data, 
manage frontends, and capture, replay, and extend tales.  The general architecture
of the platform is shown in Figure~\ref{fig:arch}. Our development philosophy follows open source principles to be consistent with our goals of research transparency, but more importantly to enable the re-use and extension of the project and encourage a community to grow around the \wt, see \url{https://github.com/whole-tale}.

\subsection{General Architecture}
At the heart of the \wt\ infrastructure lies the Metadata Management System,
which creates an abstraction layer between user-facing interfaces and the physical
location of the data. For this purpose we utilize Girder~\cite{girder}---a general purpose framework with a simple data
model and REST interface. 
%Girder supports the creation of dynamic data hierarchies that
%transparently proxy the referenced data from \emph{Assetstores}---abstractions of raw data
%storage backends. 
Using Girder, datasets can be organized into \emph{Collections}, containing \emph{Folders} and \emph{Items}. \emph{Folders}
are a hierarchically nested organizational structure that consist of other
\emph{Folders} and \emph{Items}. An \emph{Item} is the basic unit of data in the
system. \emph{Items} live beneath \emph{Folders} and contain zero or more
\emph{Files}, which represent raw data objects. Each organizational object, i.e., 
\emph{Collection}, \emph{Folder} and \emph{Item} can be annotated with metadata.
%in the form of a simple key-value pair or JSON data. 
Additionally Girder provides
models for user and group management (\emph{Users}, \emph{Groups}) and an access control model for resource management. 
Each of these objects is represented by a model with a RESTful
interface, that can be used to create, store, and retrieve persistent records
in an internal MongoDB. % via any HTTP-capable client.
As a result, data is managed entirely by reference. That is, 
the data stored in external data repositories are completely decoupled
from the data in \wt: e.g., an \emph{Item} representing an external object (e.g., an HTTP URL, or a file on a Globus endpoint) can be easily copied, renamed, and moved around
Girder's \emph{Folder} structure without performing any operations on the
actual data. This approach is advantageous as it allows \wt\ to 
scale to represent very large datasets leaving data management tasks to
external systems (e.g., Globus) and only copying the data when needed. 

% Girder can be easily extended by creating custom plugins that can
% modify or extend the server's REST API. 
The \wt\ builds upon Girder by providing plugins that:
\begin{itemize}
  \item Introduce new models for objects specific to the
    project, such as \emph{Recipe}, \emph{Image}, \emph{Tale}, \emph{Instance},
    \emph{Repository} (see Section~\ref{sec:wt_plugin} for details).
  \item Allow users to execute tasks such as building \emph{Images} from
    \emph{Recipes} and creating \emph{Instances} from \emph{Tales}.
  \item Manage transfers of streams of data from remote \emph{Repositories} into
    running \emph{Instances} (see Section~\ref{sec:dm_plugin} for details).
\end{itemize}

\subsection{The Whole Tale Workflow}
\label{sec:wt_plugin}
\begin{figure}[ht!]
\centering
  \includegraphics[trim=0in 0in 0in 0in,clip,width=0.75\columnwidth]{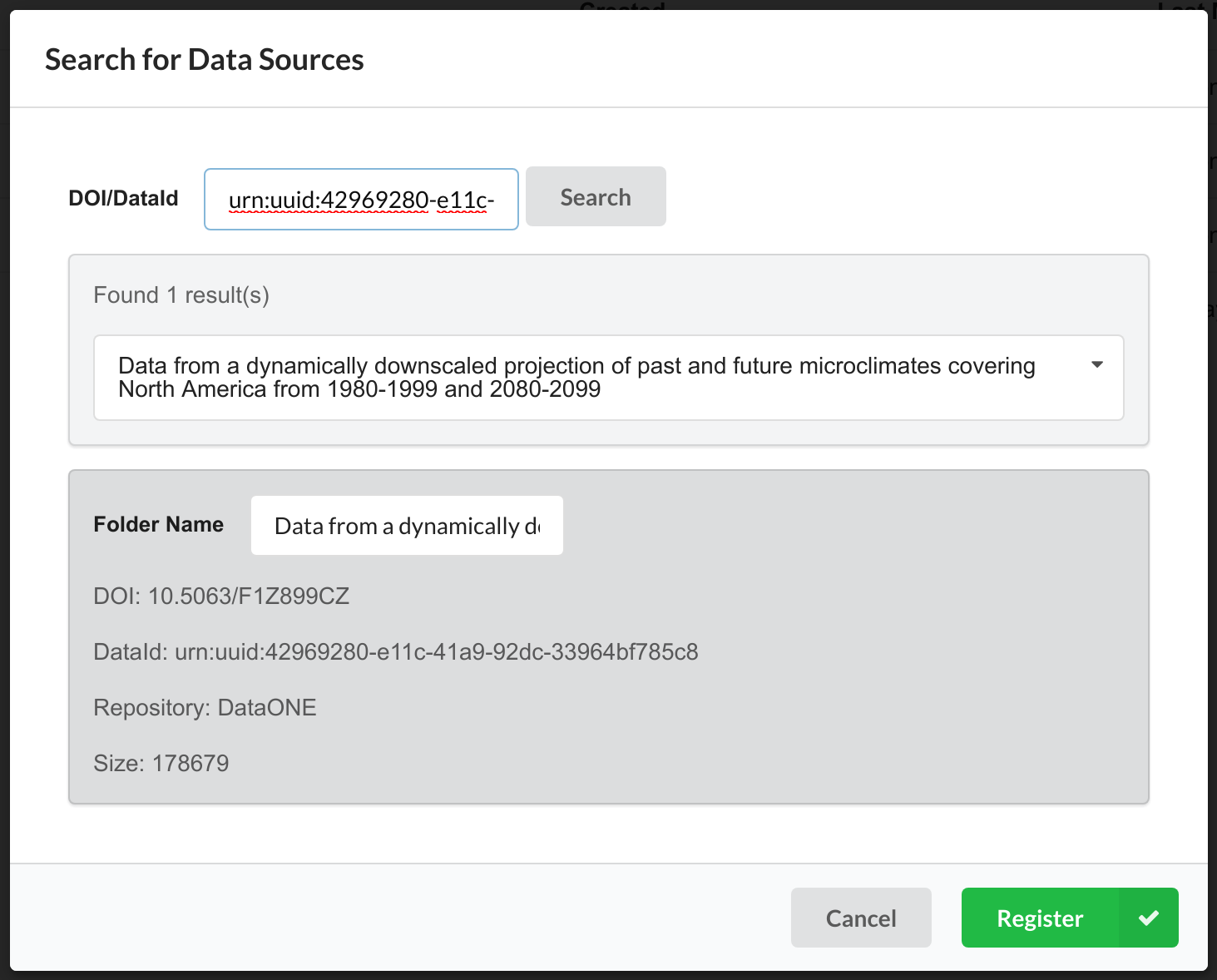}
  \caption{\wt\ data registration modal. Users provide the data unique
  identifier and register the external resources as native girder objects.
  \label{fig:reg_modal}}
\end{figure}

As mentioned in Section \ref{sec:data_ingestion}, \wt's functionality includes the reuse of published scientific data. Users may {\bf register} data located
in an external repository which \wt\ understands as native Girder objects, such as \emph{Folders} and
\emph{Items}. Registration of data, say in preparation for conducting research in \wt, is a two-step process. First, users provide a data
identifier (e.g., DOI, data provider specific UUID, URL), which is passed to the external search
engine of each supported data provider. Basic information about the dataset is obtained (e.g., name, size, provider; see
Fig.~\ref{fig:reg_modal}). To obtain this information we define a new
\emph{Repository} endpoint\footnote{\url{https://github.com/whole-tale/girder_ythub}} in Girder, which abstracts access to repository-specific interfaces.
The registration procedure starts with the creation of a
\emph{Folder} object to group all the references related to the
selected dataset (datasets may be comprised of many files and folders). 
For each of the files provided by the data provider an \emph{Item} is
created as a child object in the main \emph{Folder}. Each \emph{Item} stores the
information about the original name of the file, its location, and the protocol to
access it (e.g., HTTP, Globus, etc.).

In some cases datasets include references to other datasets. In this case, 
we create a sub-\emph{Folder} for each reference and the procedure continues
recursively. As this may be a time consuming process, users are notified of 
progress. Once the
registration is complete, users are allowed to modify the resulting data
hierarchy, i.e. rename \emph{Items}, move \emph{Folders} etc. However, these
modifications do not affect the provenance attributes of those objects (e.g., 
source repository, location, name, etc.).
It is important to note that when data is registered from a remote source
the system will provide a shallow copy of that entire 
dataset. As the user interacts with the imported dataset (e.g., in a tale),
the raw data is copied on-the fly and 
cached thereafter to create a deep copy of the data. 
At present the \wt\ supports repository access via DataONE and Globus~\cite{chard14efficient}. Additional repositories can be easily added by implementing
a simple interface that provides
the necessary information: name, size, location and access protocol;
and embedding it within the \emph{Repository} model.

The availability of the data and the fact that it can be freely composed 
into a dynamic dataset through the \emph{Folder} and \emph{Item} hierarchy,
is a necessary ingredient of the most important artifact that comes out of the
\wt\ project, which is the \emph{tale} itself. A \emph{tale} bundles a frontend and relevant data into a research environment.
The environment itself is based on a Docker image---a
lightweight, stand-alone, executable package that includes
everything needed to run a research environment for a tale.
In order to ensure that the image can be reconstructed in 
exactly the same state we require a machine
% In order to accurately define the frontend used in
% each \emph{tale}, we require a machine 
parseable description of all runtime dependencies. 
For this purpose we use a Dockerfile
as a \emph{Recipe} for constructing the environment (i.e., Docker image). 
For reproducibility purposes we treat each modification
to a given Dockerfile as a prescription for a different frontend.
% Therefore, it is natural to keep these ``recipes'' in a version controlled
% repository such as Git. 
We store these recipes using a combination of a Git repository alongside
a log of changes and represent this object as a
\emph{Recipe} in \wt. 

\begin{figure}[ht!]
\centering
  \includegraphics[trim=0in 0in 0in 0in,clip,width=0.75\columnwidth]{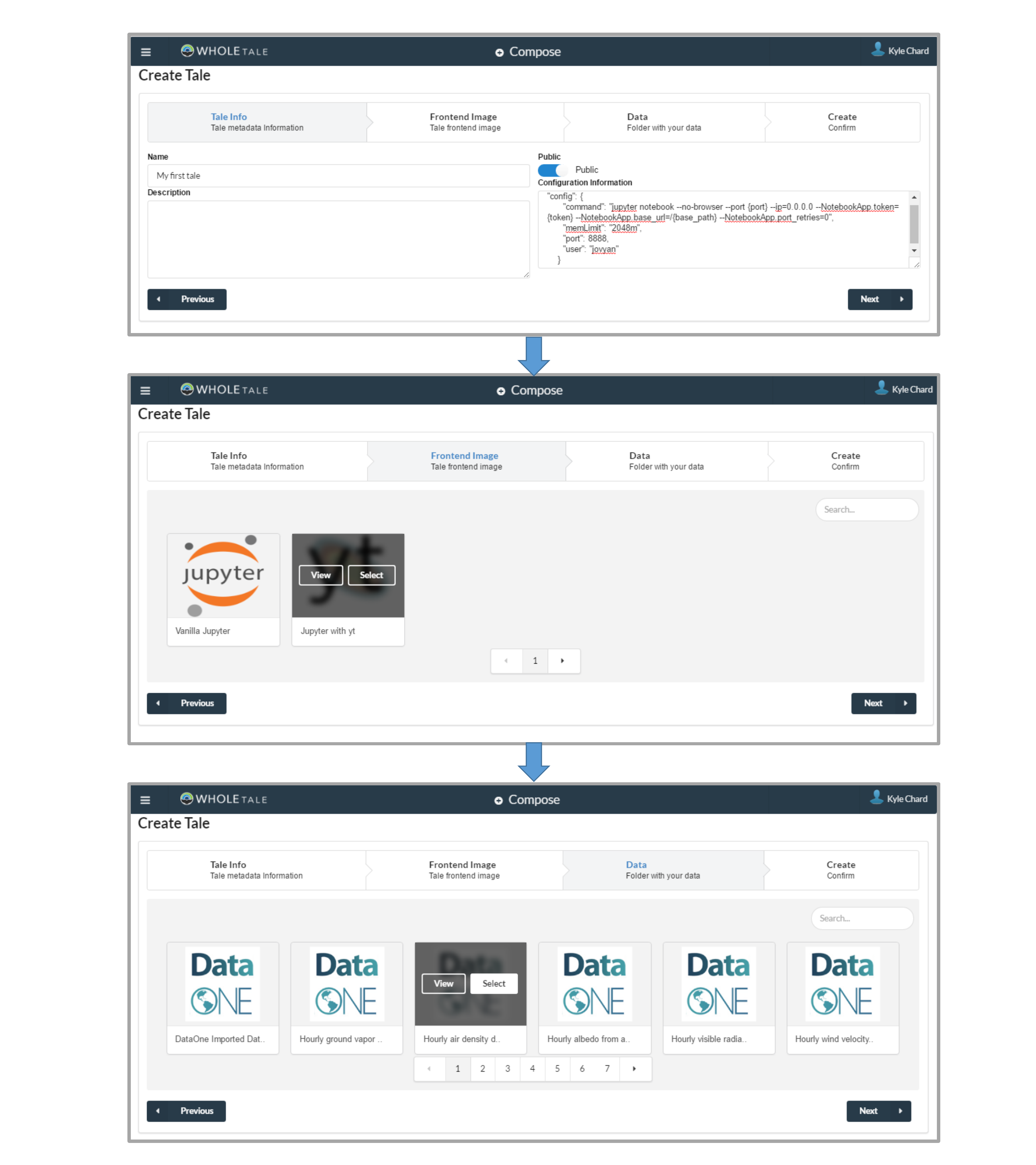}
\caption{Creating a tale via selection of a frontend and a folder containing data. \label{fig:creation}}
\end{figure}

Depending on the complexity of an image the process of building it
can be lengthy and resource consuming. We utilize a Distributed Task Queue
(Celery/ZMQ) that is integrated with Girder, to asynchronously build, track
status, and deposit Docker images in a local instance of a Docker registry---a
server application that stores and distributes Docker images. The state of a
given image is represented within the metadata management system through the
\emph{Image} model. Once a docker image is successfully built and deposited
in the Docker registry, it becomes accessible on registered \wt\ compute resources. At that point it is
available to the user as an executable research environment. The \emph{Image}, which represents the frontend, and the \emph{Folder}, that represents the data, can then be combined into a \emph{tale} (see Fig.~\ref{fig:creation}).

\subsection{Web Interface}

The \wt\ is designed to be easy to use and accessible
to a wide range of users. Its primary interface is
a web-based appolciation that allows users to manage data; create, modify, and share frontends for analyzing data; and 
create, publish, and reproduce tales by linking together datasets
and frontends. 

% The data management interface provides a comprehensive drag and drop environment 
% that can be used to discover and manage data. 
The web interface supports the standard set of file and folder operations  as one would expect in a desktop finder or file manager application (rename, move, delete, etc.). 
Files or datasets can be registered
from external data repositories (via a search workflow) or dragged and dropped from a user's desktop into the environment. Users can also view their registered files, as well as public datasets that may have been registered by other users. 
 
\begin{figure}[ht!]
\centering
  \includegraphics[trim=0in 0in 0in 0in,clip,width=0.75\columnwidth]{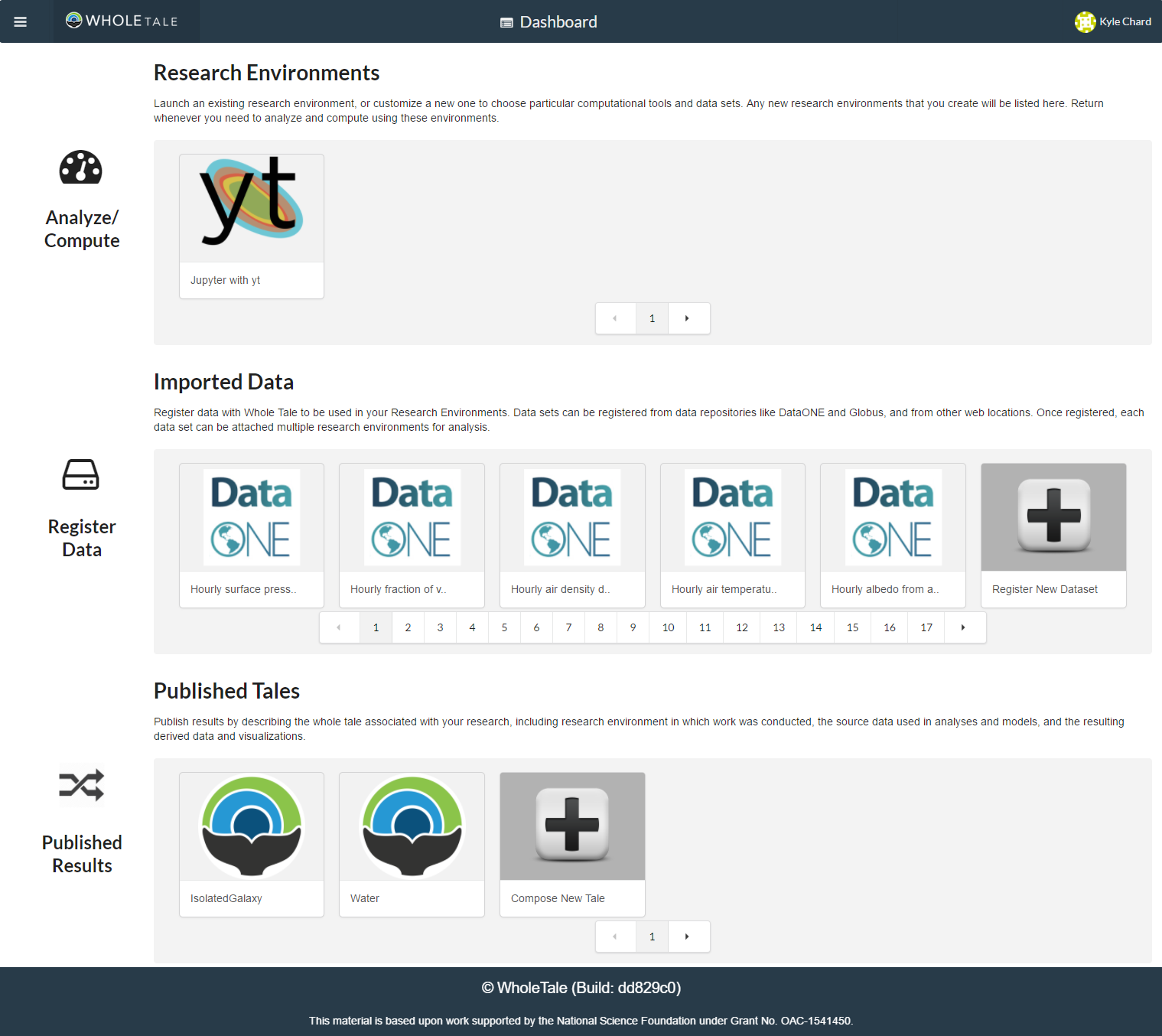}
\caption{\wt\ front page. Showing accessible frontends, datasets, and tales. Users can select an existing entity or create a new one. \label{fig:web}}
\end{figure}

The \wt\ web interface\footnote{\url{https://github.com/whole-tale/dashboard}} 
(shown in Figure~\ref{fig:web}) is implemented using the 
Ember.js open-source JavaScript web framework~\cite{emberjs}. Ember is based on the \emph{Model View View Model} (MVVM) pattern, enabling developers to create single page applications (SPAs). 
Ember also provides front end data models, which provide seamless access to Web APIs. The \wt\ interface is implemented using the Semantic UI development framework~\cite{semantic-ui}. 

% Semantic UI is powered by LESS and jQuery and has a modern design look and feel that provides intuitive and lightweight user experience. Semantic UI is also fully responsive, which enables the rapid development of interfaces for a vast range of devices.  
% Ember is 
% extremely efficient and provides a two way data binding between Javascript variables containing 
% the data and the Handlerbars templates~\cite{handlebars} that render the GUI. 

% As an analogy, Ember models are similar Django models~\cite{django}, but whereas Django models provide a Python API to a database, Ember models provide a Javascript API to a Web service.  
%Therefore, in order to implement a binding to a REST-based Web 
%service, developers simply provide a model (the data payload for the service) and the endpoint---Ember 
%provides all CRUD functionality thereafter using simple methods for get, post, save and delete, to name a few.  In the \wt\ implementation, we utilize ember models to interface to the \wt\ REST API 
%to perform operations.

\subsection{Authentication and Authorization}
We base the \wt\ authentication and authorization model on Globus Auth \cite{tuecke16auth}---a 
platform for identity and access management. 
By leveraging Globus Auth, we essentially outsource core authentication functionality
to a highly reliable service provider and need not implement
our own user management functionality (e.g., password management, user creation workflows, etc.)

Globus Auth provides a number of desirable properties for the \wt. 
First, it allows researchers to authenticate using a range of identities, 
including those common in academia (e.g., campus credentials and ORCID).
It also allows researchers to link together different identities such that
presentation of one identity enables permissions granted to any identity
in that set. 
Second, it supports standard web authentication and authorization protocols
(e.g., OpenID Connect and OAuth\,2) that simplify integration in 
\wt\ services and also provides an extensible model by which 
other related services can leverage \wt\ capabilities. Third, it provides 
an extensible delegated authorization model by which services (e.g., \wt)
can obtain delegated tokens to access other services (e.g., data repositories)
on behalf of users. Conversely, the model also allows external services 
(e.g., publishers) to obtain tokens to access \wt\ services on behalf of users. 

% We have created a Globus Auth ``App'' for the \wt\ 
We have implemented support for Globus Auth by extending Girder's OAuth plugin.
% (https://girder.readthedocs.io/en/latest/plugins.html\#oauth-login) 
This integration allows users to authenticate with \wt\ 
using any of the supported identity 
providers. \wt\ is configured to request access (``scopes'') to various resources
on behalf of users including their profile and linked identities, 
as well as being able to access other services including Globus transfer 
and MDF. 
% \wt\ follows the OAuth~2 protocol and
% receives a set of tokens for each authorized scope. These tokens are
% stored for the duration of the user's session and can be applied
% when accessing external services. 

\subsection{Data Management}
\label{sec:dm_plugin}

The \wt\ Data Management System (DMS)\footnote{\url{https://github.com/whole-tale/girder_wt_data_manager/}} is responsible for  managing the ``bits'' that make up the data used in tales. Primary data, which is data that is sourced from external services, does not, in general, come with a uniform access mechanism. Each external service is free to define its own rules and mechanisms of access. The DMS addresses this issue by providing a POSIX interface to primary data. 
This interface allows tales to act upon diverse, distributed data as if the data were local.
A secondary goal of the DMS is to provide data locality. Primary data services are assumed to be geographically distributed, as such, there is significant latency when data is accessed directly. The DMS provides an abstraction layer that hides these differences. The main components of the DMS are:

\begin{itemize}
	\item \textbf{The transfer subsystem:}
		manages the movement of data from external data providers to a storage area local to the \wt\ infrastructure. It does so through the use of transfer adapters, which are specific to each external provider.

	\item \textbf{The storage management system:} controls the use of storage space by evicting data that is considered to be unlikely to be used frequently. It acts as a data cache for external data.
	
	\item \textbf{The filesystem interface:} allows tales to access cached external data through a POSIX interface. 
\end{itemize}

From a user perspective, the process of consciously interacting with the DMS is limited to composing filesystem hierarchies to be used in tales. An initial process of ingestion described in Section~\ref{sec:wt_plugin} populates the \wt\ backend with metadata about available external data collections. Many such collections are available and navigating them in a tale, through a filesystem interface, can be difficult. Users are, therefore, allowed to freely construct specialized subsets of all the data accessible to a user and known to \wt.
% \mihael{This is technically an implementation issue stemming from the fact that Girder uses a ``exactly one parent'' model. Should it have allowed users to construct collections out of existing items, this whole composable filesystem could be replaced by a simple collection selection} 
These specialized subsets are termed ``sessions.'' Each tale is associated with a session. This association is seen by the user as a filesystem that contains the data items composing the session. The filesystem is implemented as a FUSE~\cite{fuse} layer. The filesystem is currently stored using OpenStack's Cinder block storage. 
Direct access to files on this filesystem results in data transfers from external data sources, unless the data already exists locally. A locking mechanism ensures that data corresponding to files that are in active use by a tale cannot be removed to reclaim storage space.

Maintaining local copies of all external data available to \wt\ is not feasible. Consequently, the storage management system acts as a cache and garbage collector, periodically traversing the local storage and purging data in a way that meets storage constraints as well as optimizing the latency of data access for tales. The exact optimization mechanism is flexible and involves sorting data based on an objective function that is calculated based on metadata generated by the DMS, such as usage count, usage frequency, and time of last access.

\begin{figure}[ht!]
\centering
  \includegraphics[trim=0in 3in 0in 0in,clip,width=0.75\columnwidth]{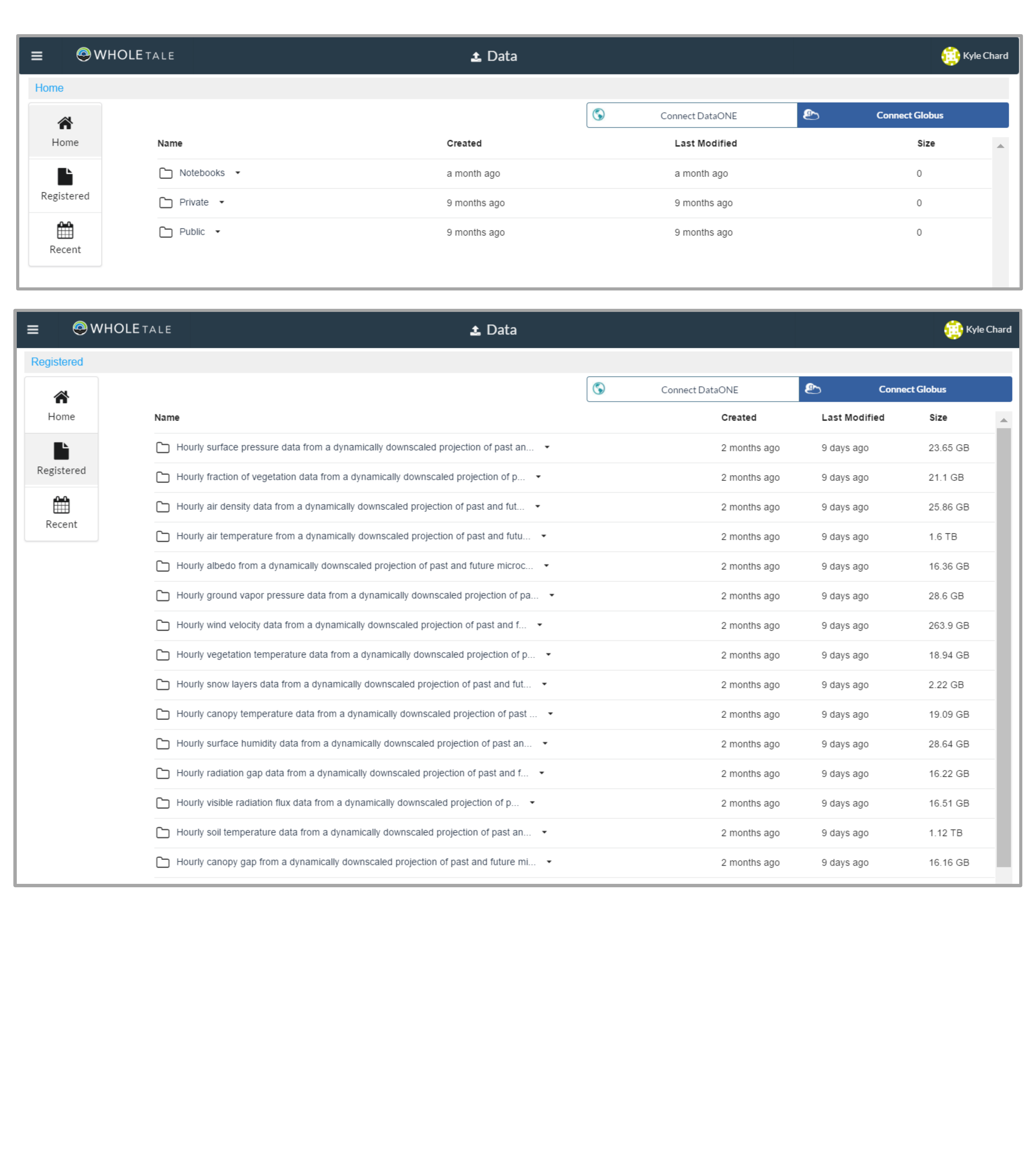}
\caption{Data management interface. \label{fig:datamanagment}}
\end{figure}

\subsection{Tale Execution and Management}

\begin{figure}[ht!]
\centering
  \includegraphics[trim=0in 3in 0in 0in,clip,width=0.75\columnwidth]{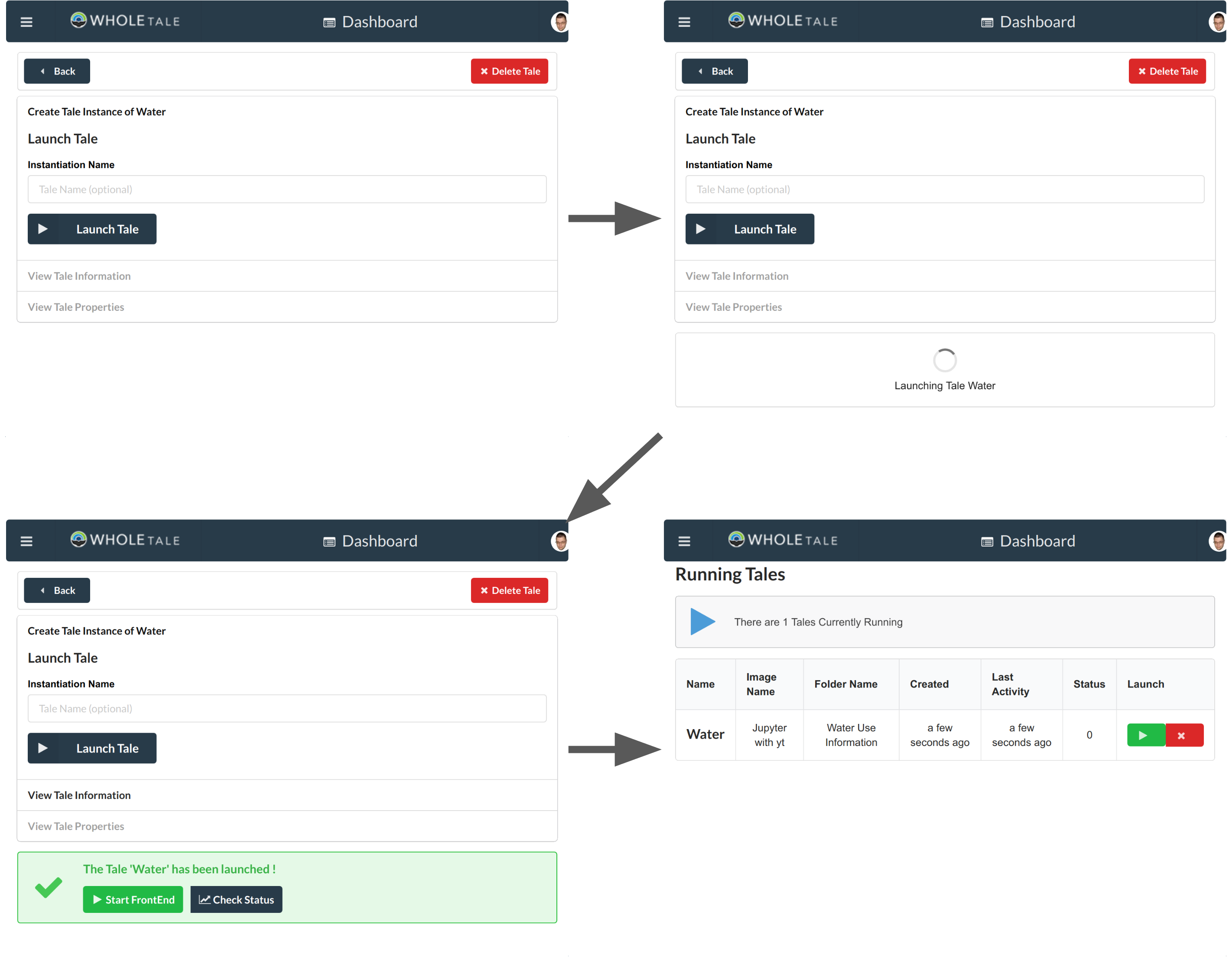}
\caption{An instantiation of a tale. \label{fig:instance}}
\end{figure}

Once a \emph{tale} has been created it can be executed (see Fig.~\ref{fig:instance}). The only requirement for execution on a specific compute resource is the availability of a Docker Engine
and two lightweight helper daemons: a reverse proxy that is responsible for routing all traffic in and out of a running tale instance (e.g., configurable-http-proxy or NGINX), and a tale management daemon (TMD)\footnote{\url{https://github.com/whole-tale/girder_volman/}} that is responsible for managing
the tale and its data dependencies. An instantiation of the tale is a multi-step process:
\begin{enumerate}
	\item A request for a \emph{tale} instance is sent from the Metadata Management System (MMS) to the TMD running on a computational cluster, along with credentials (access token).
    \item The TMD creates a docker volume using the ``local'' driver, which is basically an empty POSIX directory inside the host's filesystem.
    \item The TMD creates a docker instance using an \emph{Image} referenced by the \emph{tale} and the volume created in the previous step.
    \item The TMD creates the FUSE layer using the \emph{Folder} referenced by the \emph{tale} and mounts it in the mountpoint corresponding to the docker volume.
    \item The TMD starts the docker container and registers the internal port by which the container can be accessed with the reverse proxy.
    \item The TMD returns basic information about the container (routing path, container id, host where it is running, etc.) to the MMS.
    \item The MMS creates an \emph{Instance} model to store the information provided by the TMD and exposes it to the web interface.
\end{enumerate}

The \emph{Instance} object created during the tale instantiation is a regular RESTful object. It allows the UI to query information about running tales, and to update, suspend, or delete them. 

Tales can host \emph{any} frontend as they are based on a generic
Docker container, the only requirements of which are an open port
for user access and a mountpoint for the \wt\ FUSE filesystem. 
At present, pre-configured Jupyter and RStudio frontends are provided. These types of frontends were prioritized based on user needs and popularity.  
While users can create their own frontends, we will continue to add base
frontends to simplify use. 

\wt\ containers are executed on OpenStack virtual machines
(running Container Linux). On each VM we deploy Docker Swarm 
to manage the execution and scheduling of Docker containers
on our resource cluster. We use cloud resources at 
the National Center for Supercomputing Applications (NCSA), 
Texas Advanced Computing Center (TACC), and San Diego Supercomputer
Center (SDSC). 
There are well-established security risks of running containers
on shared infrastructure. To mitigate these risks we require that
users users upload source images to be built on-demand and
we disable intra-container communication to limit possible interference
between containers. In future work we intend to investigate 
methods for validating and certifying containers.

\subsection{Tale Representation}
Tales are defined by their execution environment, the data used, and metadata related to the tale. 
The key elements of a tale are as follows:

\textbf{Environment:} The environment captures the active components of a tale. For this purpose we rely on a Docker image and container. The tale 
maintains a reference to a Git repository (including a hash to the specific commit version). This Git repository is used as the working directory to build the docker image. 
The environment includes a name, a Git repository URL, a commit ID that references specific version of the repository, and an optional configuration object that defines specific parameters passed to Docker when running the container. 

\textbf{Data:} is represented by a set of Girder objects (folders, items, files). Each object is described with several internal descriptors (e.g., name, size, child-parent relations, uuid, creation/modification time etc.). Those that are most important to capture in a tale are the the source URL and access protocol (e.g.,  \texttt{https://website/file1}, \texttt{globus://endpoint/file1}), provider (e.g., DataONE, Globus, etc.), unique identifier (e.g., URN in DataONE, or DOI used by publication repository), and in the future an optional checksum. Given these objects are mapped to a filesystem, each object must also include its size and a POSIX compatible name. The name includes the full path (with respect to the mount point) as the tale must recreate the entire directory structure. 

\textbf{Metadata:} represents information about the tale that is
not specifically related to the environment or data. We expect
that tale metadata will grow over time based on user needs and 
tale usage.  At present, tales may include metadata that describes
the title, authors, description, icon, illustration, category (e.g., tags), publication status, as well as licensing information for all artifacts~\cite{stodden_legal_2009}.

%Running tale instances may be consider as an exploratory environment in which a researcher uses a frontend to interact with the data.
%However, an \emph{Instance} can be saved as an another \emph{tale}. In that case, the
%current state of the \emph{Instance} working environment is safely stored and becomes a
%part of new \emph{tale}.

\section{Related Work}\label{sec:relatedwork}

% The \wt\ is unlike other science gateways in some respects, however 
% it shares a number of common ideas with other related efforts. 
Scientific reproducibility is becoming an increasingly widespread
concern and stakeholders are exploring a range of 
approaches to address challenges. For example, 
data repositories now support data analysis~\cite{SciServer,TerraRef}, 
science gateways facilitate the capture of rich provenance
information~\cite{gesing15gateways},
and publishers enable verification of figures and computational 
results from within papers~\cite{shen14notebooks}, and through third party offerings~\cite{5872076, 10.1109/ICCV.2013.101}. 

Science gateways allow users to conduct (generally domain-specific) analyses that exploit
advanced computing infrastructure. They provide intuitive user interfaces that abstract
the complexities of submitting jobs via queue submission systems or
instantiating virtual computing environments for executing GUI-based tools~\cite{mclennan10hub}. Given the gateway's position at the
center of all analysis, it is possible
to capture the steps performed by users (see e.g., \cite{DBLP:journals/corr/JamesWS14}). Often these steps
are recorded in the form of workflows~\cite{goecks10galaxy}
or in other standard formats~\cite{prov-dm}.
Science gateways are generally focused on a specific domain, and 
on the analysis of data. Unlike the \wt\ they do
not provide a general model for capturing and sharing computational processes 
on arbitrary datasets and linking these artifacts with publications.

Scientific workflow systems, such as Galaxy~\cite{goecks10galaxy} and 
Kepler~\cite{CPE:CPE994}, provide
the ability for users to create flexible analysis routines 
comprising various processing steps. They typically provide extensible
interfaces via which external data can be imported for analysis. 
% Scientific workflows, by definition, capture a rich provenance trace
% in the form of the workflow description. 
While their goals overlap somewhat
with the \wt, there are significant differences between these systems.
Workflow systems prescribe a particular format and analysis model,
they therefore require researchers to modify their computational processes
to fit their model. They do not support the range of interactive 
and user-specific analyses enabled by the \wt.

As data repositories grow in size and usage there is increasing interest
in offering co-located analysis capabilities. 
Often these capabilities are offered as a set of tools
that allow users to aggregate datasets and perform simple computations. 
However, recently several data repositories have added more advanced
computing environments for processing managed data. 
For example, the Wolfram Data Repository~\cite{wolfram} provides tight coupling
with the Wolfram programming environment for analyzing and visualizing hosted data. 
The Cloud Kotta secure data enclave~\cite{babuji16kotta, babuji16secure} 
provides a co-located analysis framework that supports interactive Jupyter 
notebooks and a batch submission system for analyzing sensitive
data. These systems support only data stored within their
respective repositories. They also do not provide a model for sharing analyses 
in a standard format nor are they capable of capturing complete provenance. 

The widespread adoption of interactive programming environments
(e.g., Jupyter) have lead to countless examples of multi-user, interactive analysis environments.
For example, JupyterHub~\cite{jupyterhub}, supports multiple
Jupyter notebook instances simultaneously via execution of
notebook processes on a single server. 
Tmpnb~\cite{tmpnb} and Binder~\cite{binder} provide multi-user environments by launching
Docker containers for notebook instances. Tmpnb is used to provide
temporary notebooks for replicating analyses published 
in Nature~\cite{shen14notebooks}. 
These systems provide similar analysis capabilities to the \wt, 
however, they do not provide 
standard models for discovering and accessing data, capturing 
the computational process and the data used, or any form of linkage to publications.

Several platforms have emerged with the aim of hosting or linking digital scholarly objects to publications, including Zenodo~\cite{zenodo}, RunMyCode~\cite{stodden12runmycode}, ResearchCompendia~\cite{stodden15resaerchcompendia}, and SparseLab~\cite{sparselab}. 
These platforms typically provide a web-based location for collecting data, code, and 
other information required for verification of the published claims, with a link to 
the article or the article itself.

Publishers are providing repository services either as standalone or in support of published claims, in addition to hosting supplementary materials. Springer-Nature for example provides the figshare service~\cite{figshare}, and Elsevier provides Mendeley~\cite{mendeley}---both host digital scholarly objects such as data and code and attach unique identifiers such as digital object identifiers (DOIs) to hosted items. Other projects exist to help close similar gaps in a variety of areas. For example, the journal Image Processing Online (\url{http://ipol.im}) provides reproducible publications for the image processing community, Code Ocean provides reproducibility functionality for IEEE publications, the Madagascar project extends the reproducibility functionality described by Claerbout and Karrenbach in 1992 (\url{http://www.ahay.org}), and the WaveLab project pioneered reproducibility in signal processing (\url{http://statweb.stanford.edu/~wavelab/}), just to name a few.

There are other general efforts aimed at aggregating 
digital resources and capturing the provenance of research artifacts.
W3C PROV~\cite{w3c-prov-primer} defines a model 
for representing provenance using a data model that 
represents the entities (e.g., files), agents (e.g., people), and activities (e.g., computational processes)
associated with data processing.  
 Research Objects (RO)~\cite{bechhofer2010research}
provides a model for capturing a single unit of research including, for example,
the datasets, analysis scripts, and derived results associated 
with a paper. RO provides a formal specification for 
encoding these objects, as well as associated attribution
and provenance information. W3C PROV and ProvONE \cite{cuevas2015provone},  a PROV extension 
to link prospective and retrospective provenance \cite{mcphillips2015retrospective}, have been incorporated into DataONE \cite{cao_dataone:_2016}. 
We will be adding similar provenance support to \wt\ in the future.

\section{Summary}\label{sec:summary}

The widespread adoption of computational and data-driven science have significantly
altered the discovery lifecycle. However, the methods by which scientific results are 
published have not kept pace with the drastic changes to the underlying processes used for discovery. 
The \wt\ aims to redefine the model via
which computational and data-driven science is conducted, published, 
verified, and reproduced. The \wt\ builds upon a wide range
of efforts to support data discovery and ingestion, 
analysis using flexible frontends, scalable
computation in isolated containers, and ultimately
publication of verifiable and reproducible processes
using these artifacts. 

The \wt\ architecture consists of a set
of microservices (e.g., for data
access, persistent identifier creation, etc.) and
interoperability software that leverages, where possible, 
existing cyberinfrastructure. The resulting services
not only provide value to end users through the \wt\ 
 web interface but also to application developers through
REST APIs. 
% Its services could, for example, be leveraged
% by other science gateways to provide a variety of functions. 
Through a number of \wt\ working groups, we are actively engaging several science communities
to pilot these capabilities and evaluate their use for
enabling reproducible science.

\section*{Acknowledgments}
%\paragraph{\textbf{\textup{Acknowledgments}}} 
This research was supported by NSF grant
1541450: \emph{CC*DNI DIBBS: Merging Science and Cyberinfrastructure Pathways: The \wt}.

\section*{References}

\small
\bibliographystyle{alpha-initials-big}
%%%%%%%%%%%%%%%%%%%%%%%
\bibliography{references}

\end{document}